# A novel quad-channel 10 Gbps CMOS VCSEL array driver with integrated charge pumps

X. Huang,[a,b,**] D. Gong,[b,*] D. Guo,[a] S. Hou,[c] G. Huang,[a,*] S. Kulis,[d] C. Liu,[b] T. Liu,[b] P. Moreira,[d] A. Sánchez Rodríguez,[d] H. Sun,[a,b,**] Q. Sun,[b] J. Troska,[d] L. Xiao,[a] L. Zhang,[a,b,**] W. Zhang,[a,b,**] and J. Ye[b]

[a] *Central China Normal University,*
  *Wuhan, Hubei 430079, P.R. China*

[b] *Southern Methodist University,*
  *Dallas, TX 75275, U.S.A.*

[c] *Academia Sinica,*
  *Nangang, Taipei 11529, Taiwan,*

[d] *CERN,*
  *1211 Geneva 23, Switzerland*
  *E-mail*: dgong@smu.edu and gmhuang@mail.ccnu.edu.cn

ABSTRACT: We present a novel design and the test results of a 4-channel driver for an array of Vertical-Cavity Surface-Emitting Lasers (VCSELs). This ASIC, named cpVLAD and fabricated in a 65 nm CMOS technology, has on-chip charge pumps and is for data rates up to 10 Gbps per channel. The charge pumps are implemented to address the issue of voltage margin of the VCSEL driving stage in the applications under low temperature and harsh radiation environment. Test results indicate that cpVLAD is capable of driving VCSELs with forward voltages of up to 2.8 V using 1.2 V and 2.5 V power supplies with a power consumption of 94 mW/channel.

---

[*] Corresponding authors.

[**], visiting scholar at SMU and performed this work at SMU.

# Contents

## 1. Introduction

With developments started for the high luminosity LHC (HL-LHC) [1], optical links in high-energy physics (HEP) experiments are reaching data rates of 10 Gigabit-per-second (Gbps) per channel. The core devices for optical data transmission in HEP applications, such as the serializer/deserializer [2] and the driver for array Vertical-Cavity Surface-Emitting Lasers (VCSELs) [3][4], have been developed and demonstrated in recent years. Radiation-tolerant optical fibers [5] and VCSELs [6] have also been studied and characterized by the Versatile Link Plus group. Some research indicates that the forward voltage of VCSEL diodes may increase in high radiation environment [7], posing a problem when the driver supply voltage is 2.5 V. A low-temperature environment, which is common for the front-end readout electronics of inner tracking systems [8], aggravates this problem. Only 1.2 V and 2.5 V are available in the powering system used in the current driver development [9]. The gradual increase of the VCSEL forward voltage due to radiation will compress the voltage headroom of the driving stage and reduce the swing amplitude of the driving current to the VCSEL. In the ASIC cpVLAD [10], based on our past VCSEL driver developments [11][12], we integrate on-chip charge pumps to automatically raise the voltage to the output driving stage based on the feedback from the VCSEL forward voltage change. In this paper, we present the design and test results of the cpVLAD prototype. Section 2 describes the design. Section 3 presents the full function measurement and the test in x-ray for total ionizing dose (TID) effects. We conclude in Section 4.

## 2. Design of cpVLAD

### 2.1 Overall structure

The block diagram of the four channel VCSEL array driver, cpVLAD, is shown in Figure 1. Each channel of cpVLAD consists of a limiting amplifier (LA), an output driver, and a charge pump with feedback from the VCSEL's DC voltage level. All four charge pumps are driven by the same Voltage-Controlled Oscillator (VCO). The charge pump receives a 2.5 V power supply and a programmable clock with an amplitude of 1.2 V from the VCO and outputs a boosted voltage of up to 3.3 V for the output driver. The feedback circuit works as a low-pass filter to acquire the common-mode voltage of the output node so that the charge pump output can be automatically controlled by a closed loop. The modulation current of cpVLAD is programmable up to 14 mA. All the transistors are carefully biased to ensure that the voltage difference between any two terminals does not exceed 1.2 V.

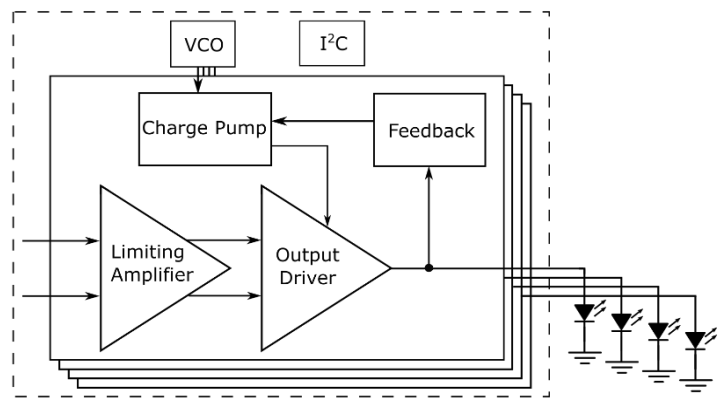

Figure 1. Block diagram of cpVLAD.

### 2.2 Limiting Amplifier

To reduce the Inter-symbol interference (ISI) jitter, the LA is designed to provide enough gain with high bandwidth. The LA receives and amplifies the input signals with various amplitudes to a saturation amplitude so that the modulation current of the output driver is independent of the input amplitude. Considering the minimum input amplitude of 100 mV, the total gain of the LA is designed to be greater than 18 dB with a total bandwidth larger than 10 GHz.

    The LA has four fully differential amplification stages [13] with an inductive shunt-peaking technique [14] to boost the bandwidth. A center-tapped symmetric spiral passive inductor is used. To fully take advantage of the on-chip inductor, we have two consecutive stages sharing one inductor [15]. The schematic of two consecutive stages with a shared inductor is shown in Figure 2.

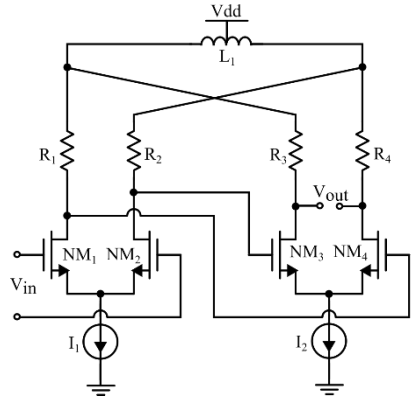

Figure 2. Schematic of the two consecutive stages of the LA.

The post-layout AC simulation result of the LA at the typical corner, the nominal power supply voltage of 1.2 V, and 27 °C is shown in Figure 3. The overall gain of the LA is 21.5 dB, and the bandwidth is 12.0 GHz.

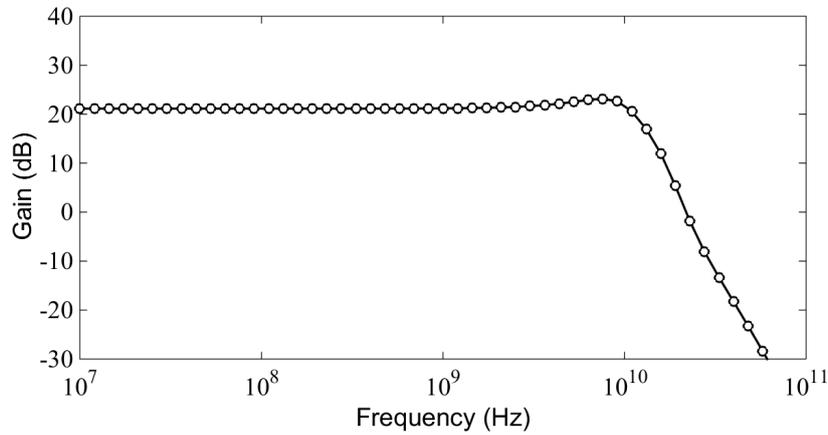

Figure 3. Gain versus frequency in the post-layout simulation of the LA.

## 2.3 Output Driver

The output stage of the VCSEL array driver provides the modulation and biasing currents to the laser diode. In order to steer the common-cathode VCSEL array directly, a single-ended output topology is chosen. The power supply to the output driver needs to be higher than the forward voltage of the VCSEL with sufficient headroom.

The schematic of the output driver is shown in figure 4. The output modulation is implemented by turning on or off the switch transistor $NM_2$ to alter the current flowing through the laser diode $D_1$. When $NM_2$ is turned off, the current $I_{tot}$ flows from the PMOS current mirror $PM_2/PM_3$ into the anode of the VCSEL. When NM2 is turned on, the current $I_{mod}$ is shunted from the total current $I_{tot}$ and the current flowing through the VCSEL is $I_{off} = I_{tot} - I_{mod}$. The currents $I_{mod}$ and $I_{off}$ can be programmed through the on-chip DACs controlled by $I^2C$. A resistor $R_d$ and a capacitor $C_d$ constitute a first-order filter to attenuate the noise of the PMOS current mirror. An active inductor consists of $R_1$ and $PM_1$ is used to improve the bandwidth of the output driver [16]. $PM_1$ also performs as protection for $NM_1$ under the power supply Voltage of the Output Driver (VDDOD). The capacitor $C_f$ works as a feed-forward capacitance path to further boost the bandwidth of the output driver [17].

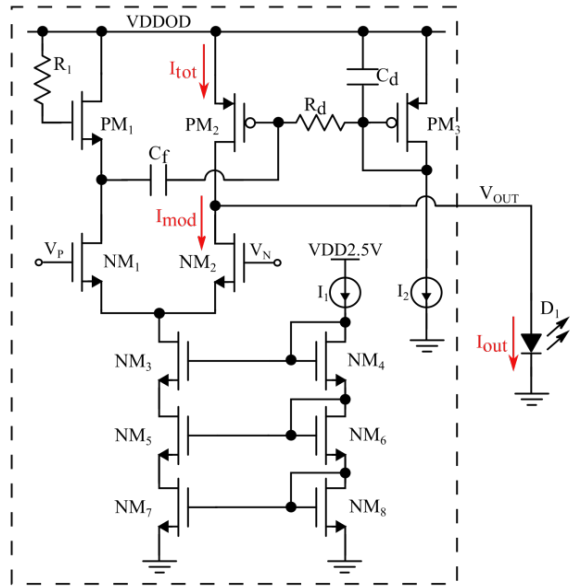

Figure 4. Schematic of the output driver.

The AC simulation result of the output driver at the typical process corner, VDDOD = 2.5 V, and 27 °C is shown in figure 5. The bandwidth reaches 20.25 GHz.

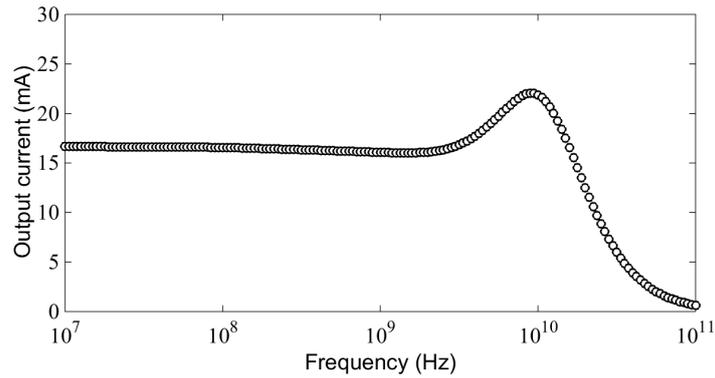

Figure 5. Output current versus frequency in the post-layout simulation of the output driver.

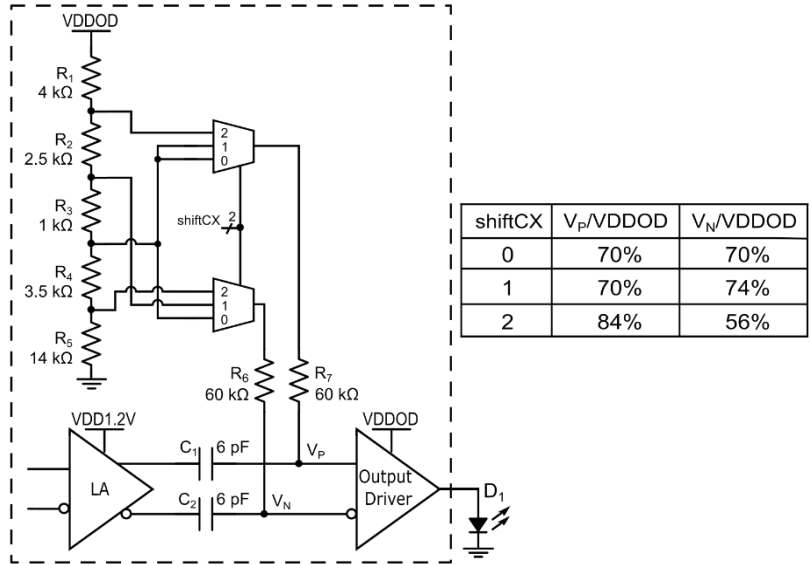

Figure 6. Schematic of the AC coupler

Because the power supply of the output driver is much higher than that of the LA, the optimal common mode voltage between the LA and the output driver does not match well. We use an AC coupling to solve this problem. The low cutoff frequency of the AC coupling circuit is 407 kHz, low enough for 10 Gbps optical data transmission. The input CM voltage of the output driver is biased by a resistor division network and two multiplexers implemented in transmission gates, shown in Figure 6. The multiplexers are controlled by two register bits shiftCX in I$^2$C. When shiftCX is 0, $V_P$ and $V_N$ are both biased to 70% of VDDOD. This is the nominal operation condition. When shiftCX is 1, $V_P$ is biased to 70% of VDDOD and $V_N$ is biased to 74% of VDDOD. This option is to compensate for the asymmetry structure of the output driver. By tuning the crossing point of the output eye diagram, the Duty Cycle Distortion (DCD) may be reduced. When shiftCX is 2, $V_P$ is biased to 84% of VDDOD and $V_N$ is biased to 56% of VDDOD. In this option, the right branch (NM$_2$ in figure 5) of the output driver is turned off and I$_{tot}$ flows into the VCSEL diode. This option provides the feature to measure the Light-Current (L-I) curve of the VCSEL diode when it is wire-bonded to the driver IC.

The output driver was simulated together with the LA and with full parasitic post-layout extractions at various Processes, Voltages, and Temperatures (PVTs). A regular 1.6 V forward voltage in the VCSEL model is used in the simulation. Figure 7 shows the output eye diagram at the typical process corner, 1.2 V (for the LA) or 2.5 V (for the output driver without the charge pump function), and at 27 °C with the default setting. The modulation current of the eye diagram is 6 mA. The simulated Inter-Symbol Interference (ISI) jitter is 4 ps (peak-peak).

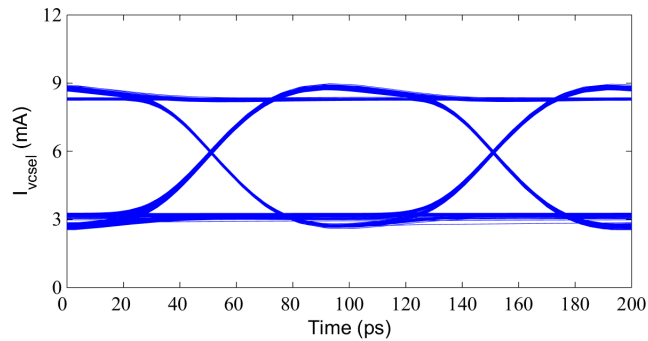

## 2.4 Charge Pump

The charge pump produces a voltage higher than 2.5 V to gain sufficient headroom for the output driver when the forward voltage of the VCSEL increases in low temperature and radiation environment.

Figure 8(a) shows the schematic of the charge pump [18]. $C_1$ and $C_2$ are the pumping capacitors, and $C_L$ is the load capacitor. The cross-connected MOS transistors ($NM_1$, $NM_2$, $PM_1$, and $PM_2$) function as switches controlled by the AC coupled clocks. The NMOS transistors $NM_1$ and $NM_2$ are utilized to charge $C_1$ and $C_2$. The PMOS transistors $PM_1$ and $PM_2$ operate to transfer the charge from $C_1$ and $C_2$ to $C_L$. The symmetrical charge-discharge branches are sequentially controlled by two complementary 800 MHz clocks, CLK and $\overline{\text{CLK}}$, from an on-chip VCO. Figure 8(b) and 8(c) demonstrate the process of the charge pump. In Figure 8(b), CLK is high and $\overline{\text{CLK}}$ is low. $NM_1$ and $PM_2$ are turned on, while $NM_2$ and $PM_1$ are turned off. Capacitor $C_1$ is now charged by the 2.5 V power supply. Meanwhile, charge in capacitor $C_2$ is diverted to $C_L$ to promote the output voltage. Figure 8(c) shows the complementary case, when CLK is low and $\overline{\text{CLK}}$ is high, $C_2$ is charged by the 2.5 V power supply and $C_1$ works as the output source. The clock signals offer a DC potential $V_{CLK}$ of 1.2 V at each discharging period, and the output voltage is boosted based on $V_{CLK}$. The voltage drop of each transistor is under $V_{CLK}$ so that standard transistors can be adopted to reduce voltage threshold and parasitic capacitance and to increase current drive capability.

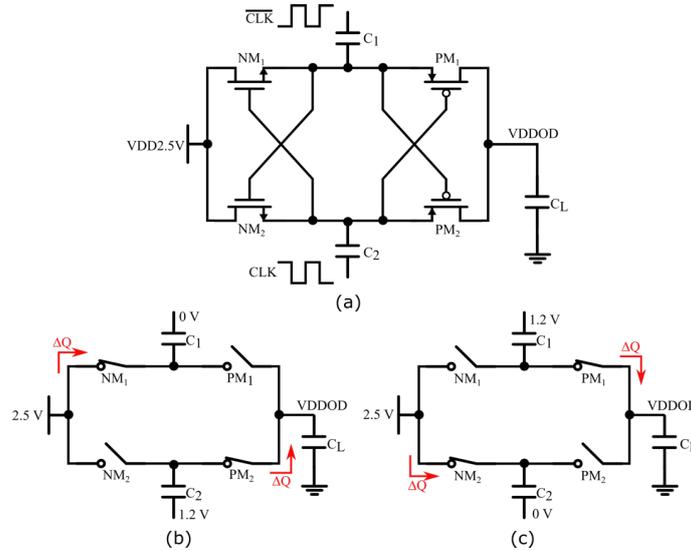

Figure 8. Schematic of the charge pump (a), simplified schematic when CLK is high (b), and simplified schematic when CLK is low (c).

Furthermore, a negative feedback loop is used to automatically adjust the voltage VDDOD to the output driver to ensure that the output driver operates in a proper state. The schematic of the feedback circuit is depicted in figure 9. A first-order RC filter, which consists of $C_1$ and $R_3$, senses the forward DC voltage of the VCSEL. A feedback path including an operational amplifier (OPA) is introduced to keep the difference between the charge pump output voltage VDDOD and the VCSEL forward voltage to be constant at $\Delta V = I_1 R_2$. The OPA controls the current shunt branch, which consists of $R_1$ (879 Ω) and $PM_1$ (on-state resistance $R_{on}$ from 1 Ω to 1.65 Ω), to balance the output current flowing through the output driver.

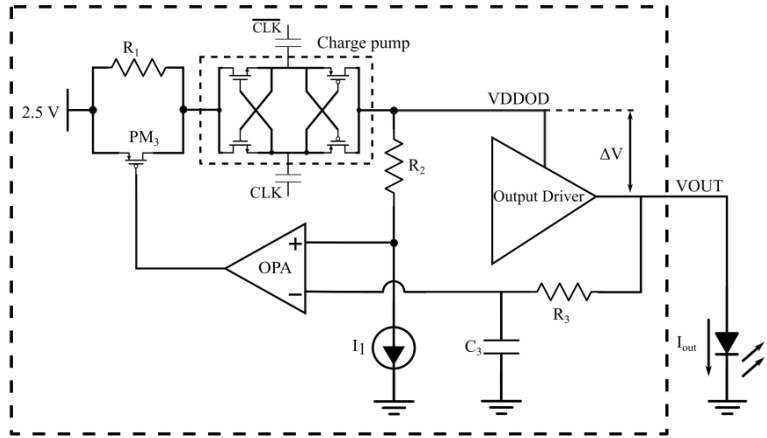

Figure 9. Schematic of the feedback circuit in the charge pump.

The start-up process of the charge pump in the post-layout simulation at the load current of 16 mA is shown in figure 10. After the charge pump is powered on for 30 ns, the output voltage of the charge pump stabilizes around 3.2 V with a peak-to-peak ripple of 60 mV under the typical PVT. The voltage ripple can be decreased by enlarging the load capacitor or by increasing the clock frequency. An optimal combination of $C_L$ (155.4 pF) and VCO frequency (800 MHz) is chosen based on the consideration of area and process performance. The layout of the charge pump is designed to be as symmetrical as possible so that the same charge can be transferred per half cycle to produce the identical output voltage. The charge pump module occupies an area of $0.2 \times 1.0$ mm$^2$.

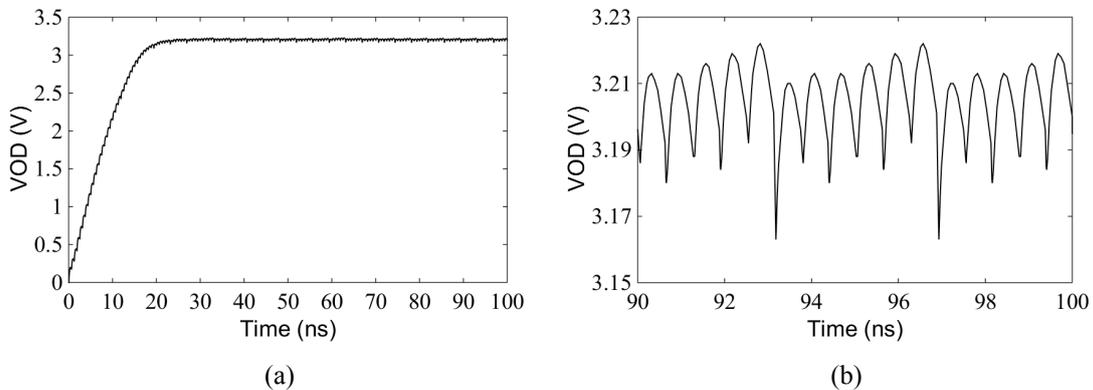

(a)          (b)

Figure 10. Start-up process of the charge pump (a) and voltage ripple of the charge pump output (b) in the post-layout simulation.

Figure 11 is the simulated efficiency of the charge pump under different load currents. When the load current is a typical operating current of 16 mA, the power efficiency of the charge pump is 70.5 %.

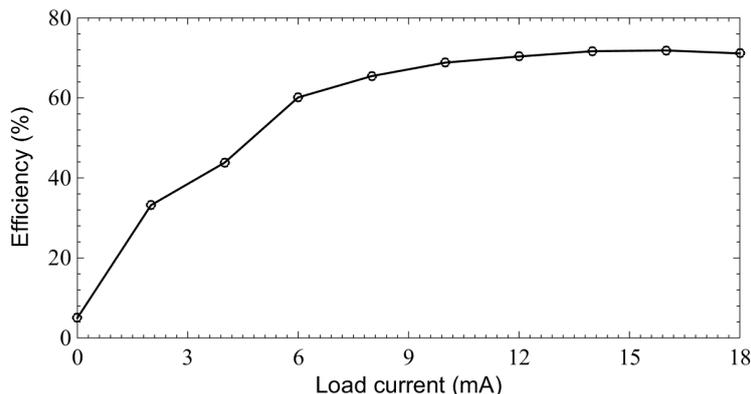
Figure 11. Simulated efficiency of the charge pump under different load currents.

**2.5 Auxiliary circuits and overall layout**

A VCO generates two complementary clock signals for all charge pumps. The VCO is a 4-stage ring oscillator. The frequency of the VCO is programmable. The frequency adjustment is achieved by tuning the operation current of the inverters to change the delay time. The tuning range of the VCO is from 600 MHz to 900 MHz. The I$^2$C module is implemented with triple modular redundancy to configure the internal registers.

The cpVLAD prototype is fabricated in a 65 nm CMOS technology. The differential high-speed input signal paths are in the thick metal layer to reduce the parasitic resistance. The output driver circuits are located close to the output pads to reduce parasitic capacitance. To reduce the voltage noise from the power supply, all unused area is filled with custom decoupling capacitors. The charge pump and the decoupling capacitor of the power supply of each channel are independent to reduce crosstalk. The power supplying grids are implemented by interweaving two different thick metal layers to ensure the current margin and the solid contact between the power pads and the internal circuits. The area of cpVLAD is 1.835 mm x 1.635 mm.

**3. Test results**

The cpVLAD prototype has been characterized in both the electrical test and the optical test using two separate boards. For the electrical test, cpVLAD was wire-bonded on the board as shown in Figure 12(a), and the output was connected to an oscilloscope through an external bias-tee with an adjustable DC voltage to emulate the VCSEL forward voltage. For the optical test, a commercial 25 Gbps 850 nm VCSEL array (Philips, ULM850-25-TT-N0104U) was wire-bonded to cpVLAD, and the optical signals from the VCSEL array were coupled to optical fibers through a commercial coupler. It should be noted that we use a 25 Gbps VCSEL array with the forward voltage as high as 2.3 V to demonstrate the charge pump capability. Figure 12(b) shows a photograph of the optical assembly, including the coupling component, the cpVLAD ASIC, and the VCSEL array underneath the coupler. During the tests, a pattern generator (Picosecond Lab, Model SDG 12070) provided the 10 Gbps differential Pseudo-Random Binary Sequence (PRBS) signals with different amplitudes. Two patterns, $2^7$-1 and $2^{31}$-1, were both used in the tests. An oscilloscope (Tektronix DSA72004 for electrical signals or Tektronix TDS8000B for optical signals) was used to capture the output eye diagrams. Two DC power supplies (Agilent E3641A) provided 1.2 V and 2.5 V power supply for the test board, respectively. A source-meter unit (Keithley, Model 2410) was used to emulate the VCSEL forward voltage. The source-meter also performed as a current source or sink. Figure 13(a) and Figure 13(b) show the block diagram of electrical test and optical test respectively. Figure 13(c) shows a photograph of the test setup.

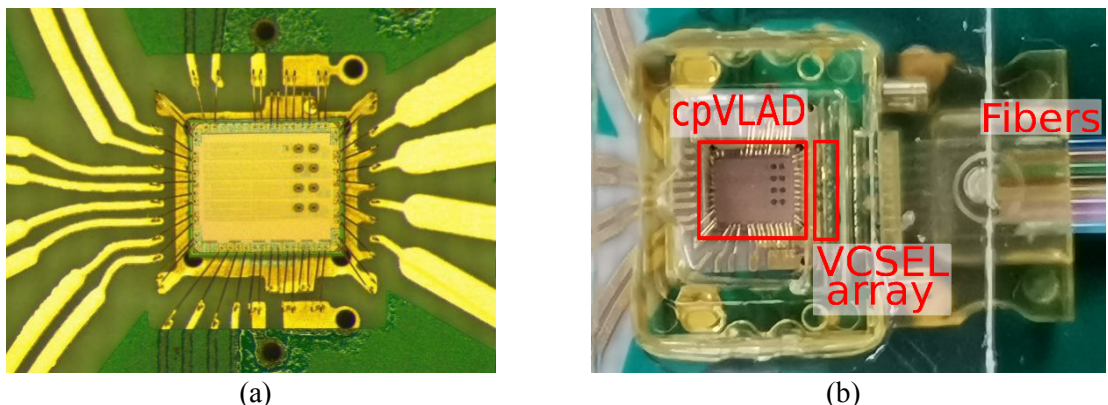

(a)            (b)

Figure 12. Photographs of cpVLAD on the electrical test board (a) and the optical assembly on the optical test board (b).

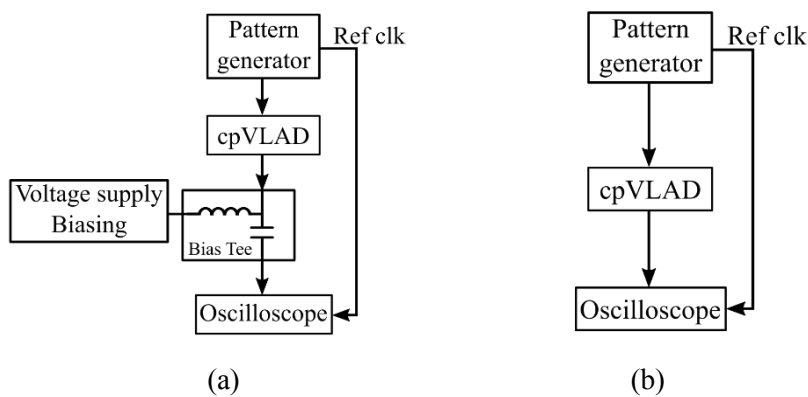

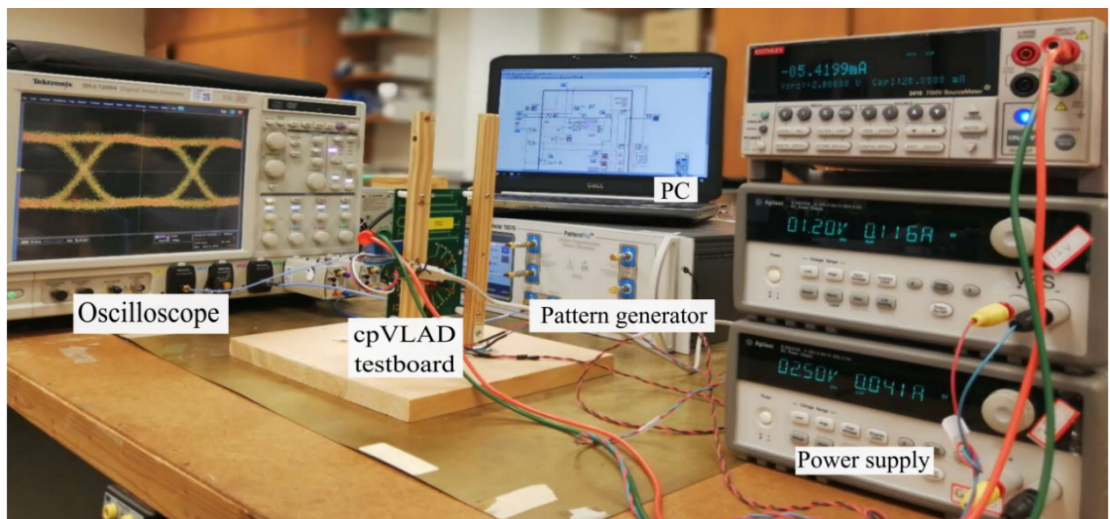

(c)

Figure 13. Block diagram of electrical test (a) and optical test (b) and photograph of the test setup (c).

The input signal level achieving the full output swing is the input sensitivity. The input signal sensitivity was measured with an electrical test board. The input amplitude varied from 50 mV to 1.2 V. The output modulation amplitude versus the input amplitude is plotted in Figure 14. The output modulation amplitude is independent of the input signal amplitude when the input amplitude is larger than 80 mV, illustrating that we achieve the designed sensitivity of 100 mV. A 200 mV input amplitude was chosen for the following measurements. The simulated results at the nominal voltages of 1.2 V for the LA and 2.5 V for the output driver, 27 °C, and different process corners (FF, TT, and SS) are also shown in the figure. The measured results match the simulated TT and SS corners.

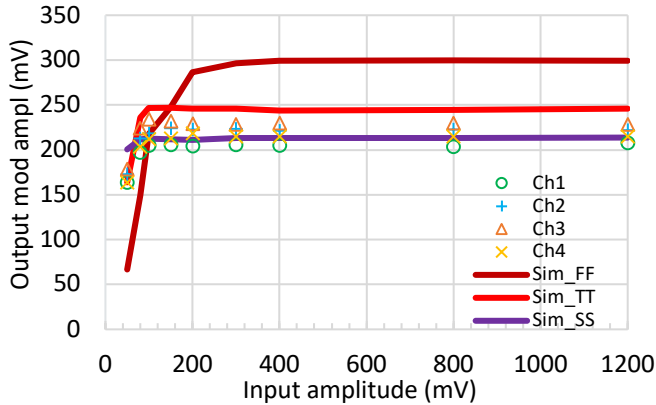

Figure 14. Input sensitivity.

In order to find the ranges of $I_{mod}$ and $I_{off}$, in which cpVLAD works, $I_{mod}$ and $I_{off}$ were scanned in both electrical and optical tests. The bias-tee current (i.e., the bias current of VCSEL) and the output modulation current were measured by sweeping the $I_{mod}$ and $I_{off}$ settings in the electrical test. The Average Optical Power (AOP) and the Optical Modulation Amplitude (OMA) were measured in the optical test. Figure 15 is the electrical output bias current (a), the electrical output modulation current (b), the AOP (c), and the OMA (d) versus the $I_{mod}$ and $I_{off}$. The OMA is not measured when $I_{mod}$ is 0, shown as white in the first column of Figure 15(d). Comparing the electrical and the optical results, we can see that the electrical output bias current and the AOP follow the same trend and correlate to both $I_{mod}$ and $I_{off}$, as we expected. The electrical output modulation current correlates only to the $I_{mod}$ but hardly the $I_{off}$. However, the optical OMA has a smaller $I_{mod}$ range than the electrical output modulation current. This reflects the fact that as the current increases, the efficiency of the VCSEL decreases.

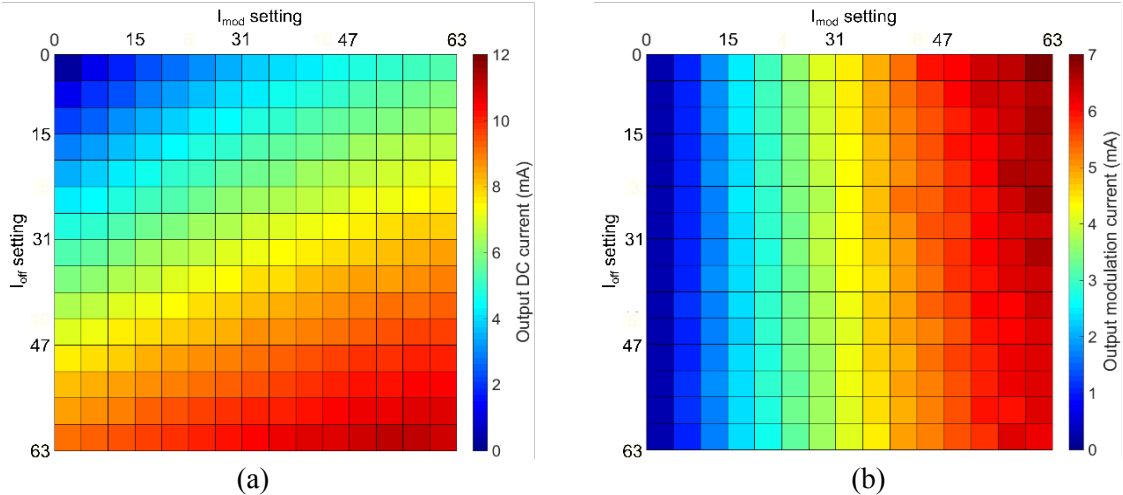

(a)  (b)

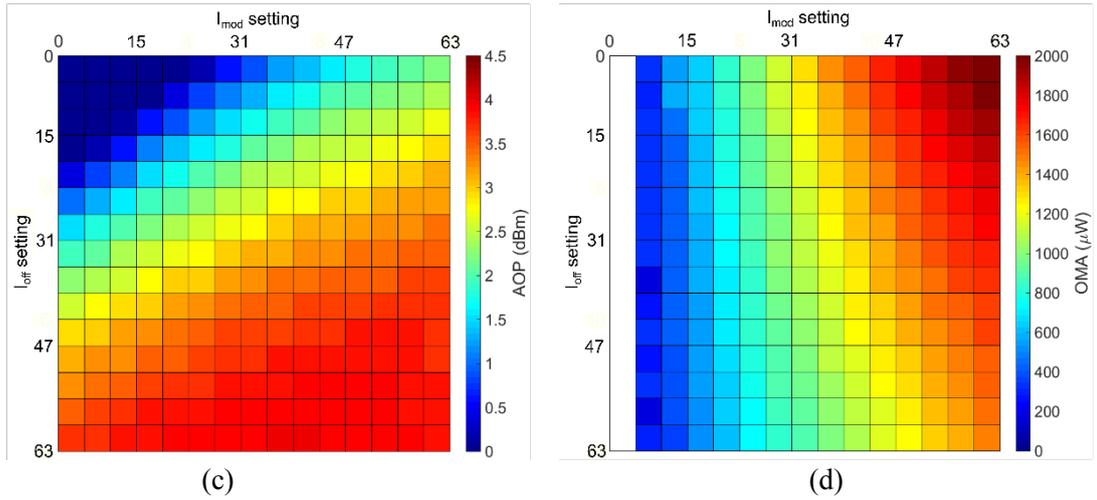

Figure 15. Output bias current versus the $I_{mod}$ and the $I_{off}$ (a), output modulation current versus the $I_{mod}$ and the $I_{off}$ (b) in the electrical test, AOP versus the $I_{mod}$ and the $I_{off}$ (c), and OMA versus the $I_{mod}$ and the $I_{off}$ (d) in the optical test.

The eye diagram and DCD adjustment feature were verified at 10 Gbps with an input amplitude of 200 mV in the electrical and optical tests. Figure 16(a) and 16(c) show the electrical and optical eye diagrams when shiftCX = 0, respectively. Figure 16(b) and 16(d) respectively show the electrical and optical eye diagrams when shiftCX = 0. We can see that the crossing point of the eye diagram moves with different shiftCX values. More steps will be added in the future to fine-tune the DCD to the optimal point. Figure 16(b) shows the 10 Gbps electrical eye diagram of channel 4 with a 6.3 mA modulation current and a 2.1 mA offset current. The total jitter is 24.4 ps (peak-to-peak) at the Bit Error Rate (BER) below $1\times10^{-12}$ with a random jitter of 0.91 ps (RMS). Figure 16(d) shows the 10 Gbps optical eye diagram of Channel 4 with the same setting as in the electrical test. The AOP is 3.41 dBm, and the OMA is 960.97 μW. The optical eye diagram passed the 10 Gbps eye mask [19] test.

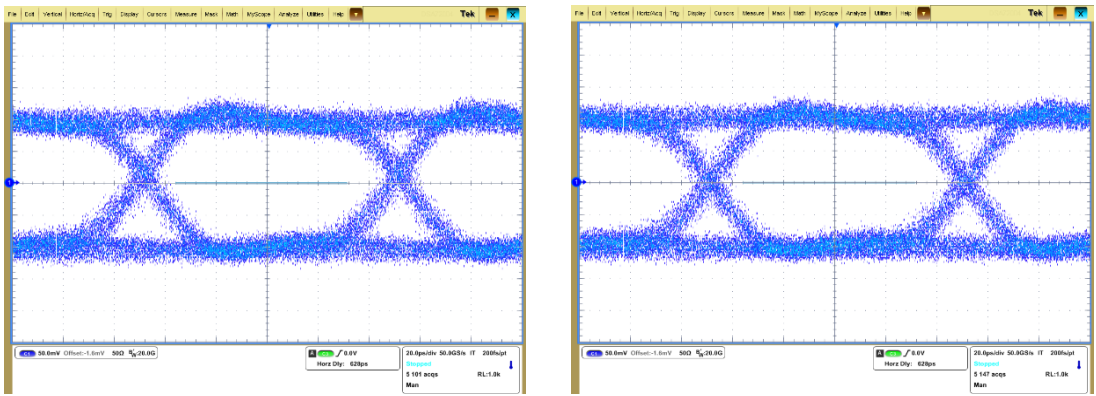

(a)            (b)

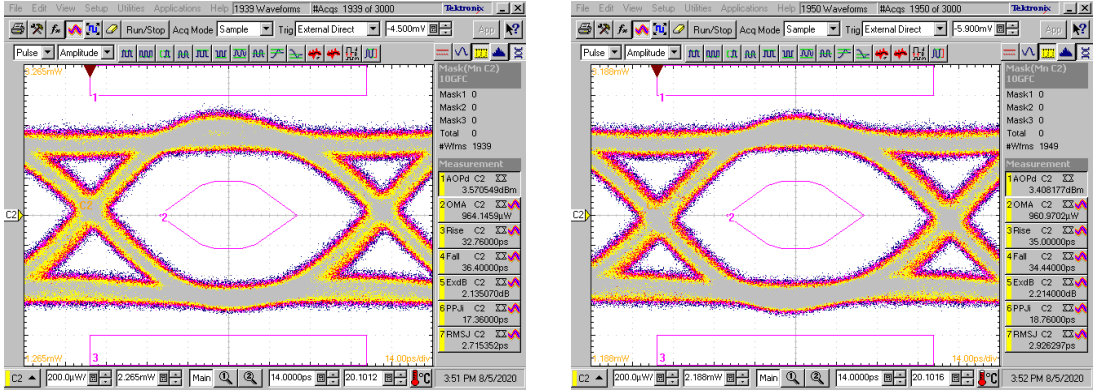

|     |     |
| :---: | :---: |
| (c) | (d) |

Figure 16. Electrical eye diagrams when shiftCX = 0 (a), when shiftCX = 1 (b). Optical eye diagrams when shiftCX = 0 (c), when shiftCX = 1 (d).

As mentioned in Section 2.3, cpVLAD features the capability to measure the L-I curve of the VCSEL diode. Since the current flowing through the VCSEL diode ($I_{out}$) cannot be accessed on the optical test board, we consider to use the increment current of the 2.5 V power supply ($\Delta I_{2.5V} = I_{2.5V} - I_{2.5V(Itot=0)}$) to stand for $I_{out}$. It should be noted that a small current of 2.5 V (i.e., $I_{2.5V(Itot=0)}$) is consumed by the output driver and the charge pump when $I_{tot}$ is set to be 0. $I_{out}$ and $\Delta I_{2.5V}$ are measured on the electrical test board and shown in Figure 17(a). The charge pump is turned off in the test. The forward voltage ($V_F$) of the VCSEL diode is set to 2.0 V, the forward voltage of a typical VCSEL diode, or 2.3 V, the $V_F$ of the VCSEL diode installed on the optical test board. An offset of 1-2 mA exists in $I_{out}$ when $\Delta I_{2V5}$ equals to 0 mA because the nonzero minimum current exists in DACs to protect the transistors used in the DACs. Except for the offset, $\Delta I_{2.5V}$ approximately equals to $I_{out}$. When the forward voltage is 2.0 V, the maximum of $I_{out}$ exceeds 17 mA, high enough to cover most VCSEL diodes operating at 10 Gbps. When the forward voltage is 2.3 V, the maximum of $I_{out}$ is less than but close to 10 mA, due to the voltage headroom problem that we hope to use the charge pump in the cpVLAD to solve. The L-I curve measurement capability is confirmed on the optical board. AOP and $\Delta I_{2.5V}$ are measured by sweeping $I_{tot}$ setting from 0 to 63. $I_{mod}$ is set to 0 and shiftCX is set to 2. The charge pump is turned off. Figure 17(b) shows the measured L-I curve of the VCSEL diode.

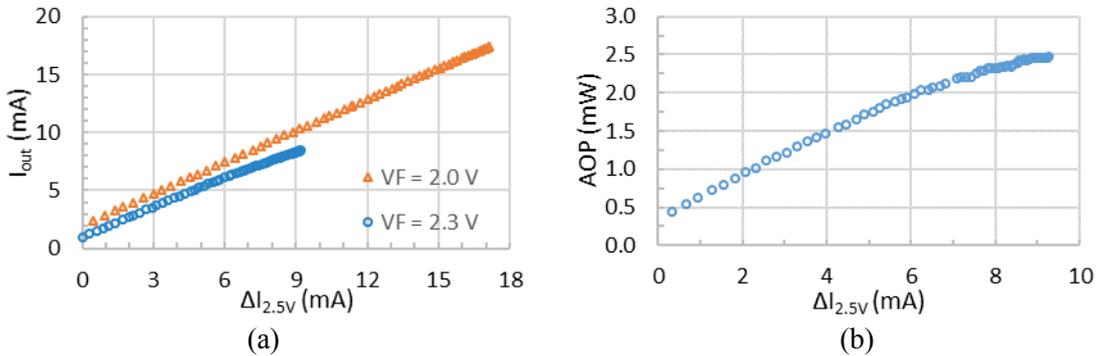

|     |     |
| :---: | :---: |
| (a) | (b) |

Figure 17. Current flowing through bias-tee (a) and AOP (b) versus the current increment of the 2.5 V power supply.

The charge pump function was verified in the electrical test by increasing the DC voltage of the external bias-tee structure. This DC voltage emulates the VCSEL forward voltage and was scanned from 1.6 V to 3.2 V (charge pump on) during the test. The output modulation amplitude versus the emulated VCSEL forward voltage is shown in Figure 18. All the channels with charge pumps on had stable modulation outputs up to a 2.8 V forward voltage. The simulated results at different process corners (FF, TT, and SS), the nominal power supply voltages of 1.2 V (for the LA) and 2.5 V (for the charge pumps), and 27 ºC are also shown in the figure. The measured results again match the simulated results of the TT and SS corners.

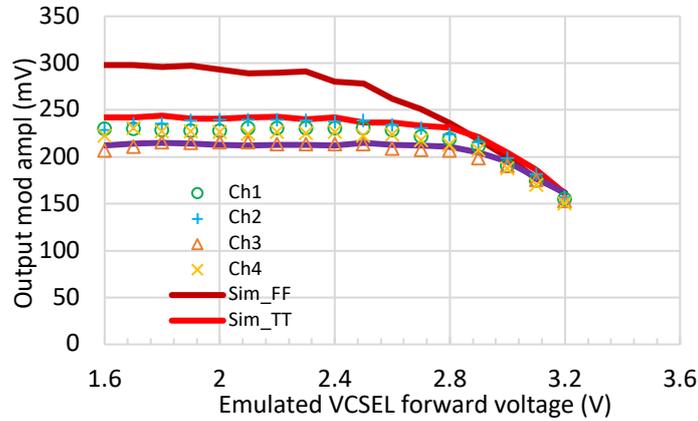

Figure 18. Output modulation amplitude versus emulated VCSEL forward voltage.

Crosstalk among channels was measured. There is no significant difference in total jitter between the two scenarios that only one channel was turned on and that all four channels were turned on, meaning that the crosstalk was negligible.

An electrical board was tested over a temperature range from -35 to +60 ºC in a climate chamber (CTS, series T) at CERN. The voltage levels of the bias-tee were set from 1.8 V to 3.3 V in a step of 0.1 V. The currents of the 2.5 V and 1.2 V power supplies and the current passing through the bias-tee were monitored at 8 temperatures (-35, -25, -10, 0, 10, 25, 45, and 65 ºC). Eye diagrams were measured at -35, 25, and 65 ºC. cpVLAD works over the whole required ambient temperatures. The eye diagrams at two extreme temperatures are shown in Figure 19. Compared with room temperature, the current changes of the 1.2 V power supply, the 2.5 V power supply, and the bias-tee over the whole temperature range are within 7.7%, 14.3%, and 13.8%, respectively.

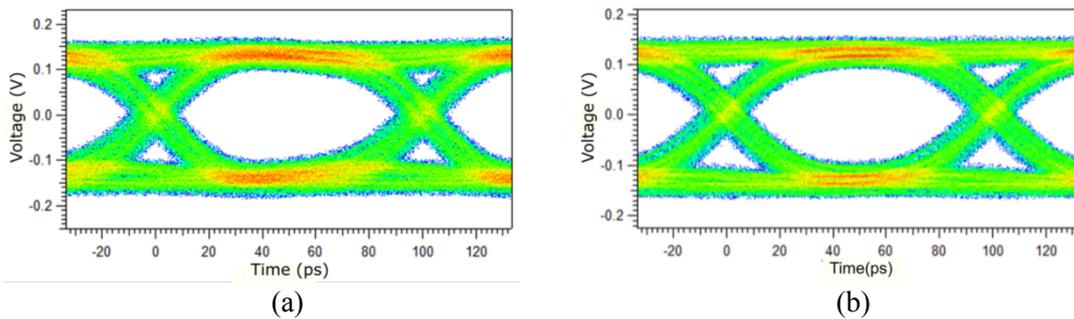

(a)                    (b)

Figure 19. Eye diagrams at -35 ºC (a) and +65 ºC (b) at the VCSEL forward voltage of 2.8 V.

The irradiation tests were carried out on an optical test board and an electrical board in the x-ray irradiation facility called ObeliX at CERN. The electrical test board was irradiated to 9 MGy for 69 hours with a dose rate of 130 kGy/h. The optical board was irradiated to 6.3 MGy for 90 hours at a dose rate of 70 kGy/h. During the tests, the chips were powered up. For the optical board, the currents of the power supplies and the eye diagrams were recorded. For the electrical board, the currents of the power supplies (1.2 V and 2.5 V), the bias-tee current, the modulation amplitude were monitored. For the optical board, the OMA and the jitter versus the total ionizing dose are shown in Figure 20. As can be seen in the figure, the OMA and the jitter have no significant change. The currents of 1.2 V and 2.5 V power supplies decreased by 7% and 3.3%, respectively, within the specification at 1 MGy. For the electrical board, the electrical output modulation and the electrical bias current maintain stable. The currents of 1.2 V and 2.5 V power supplies decreased by 7% and 3.3%, respectively, within the specification at 1 MGy. A Single Effect Upset (SEU) test will be carried out in the future.

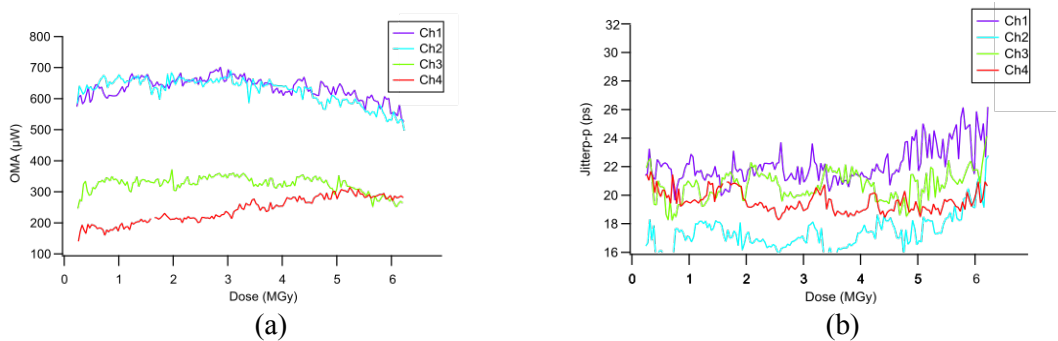

Figure 20. Optical eye height (a) and optical jitter (b) versus the total dose.

## 4. Conclusion

This paper describes the design and the performance of a 4 × 10 Gbps VCSEL array driver, cpVLAD, which is fabricated in a 65 nm CMOS process. The die of cpVLAD is $1.835 \times 1.635$ mm$^2$. To address the issue of the voltage margin for the VCSELs resulting from low temperature and radiation in high-energy physics experiments, charge pumps are integrated into each channel of cpVLAD. The test results demonstrate that cpVLAD meets the design goals that allow VCSEL to operate with a forward voltage up to 2.8 V. The power consumption is 94 mW/channel.

## Acknowledgments

This work is supported by SMU's Dedman Dean's Research Council Grant, and the National Science Council in Taiwan. We want to thank James Kierstead from Brookhaven National Laboratory for the help in the irradiation test.